\documentclass[%
 reprint,
 amsmath,amssymb,
 aps,
]{revtex4-1}

\usepackage{graphicx}
\usepackage{dcolumn}
\usepackage{bm}
\usepackage{color}
\usepackage{soul}

\newcommand{\eqLabel}[1]{\label{eq:#1}}
\newcommand{\figLabel}[1]{\label{fig:#1}}

\newcommand{\fref}[1]{Fig.~\ref{fig:#1}}
\newcommand{\eref}[1]{Eq.~(\ref{eq:#1})}
\newcommand{\sref}[1]{Sec.~\ref{sec:#1}}

\newcommand{\bv}[1]{\mathbf{#1}}
\newcommand{\lr}[1]{\left( #1 \right)}
\newcommand{\LR}[1]{\left[ #1 \right]}
\newcommand{\cLR}[1]{\left\{ #1 \right\}}
\newcommand{\qd}[1]{\left\langle #1 \right\rangle}
\newcommand{\abs}[1]{\left| #1 \right|}

\newcommand{\mat}[1]{ \underline{ \underline{ \bv{#1} } } }

\newcommand{\unitVector}[1]{\bv{e}_{#1} }

\newcommand{\partialD}[2]{\frac{\partial #1}{\partial #2} }

\newcommand{\pressure}{\mathcal{P}}
\newcommand{\stress}{\mathcal{S}}
\newcommand{\viscosity}{\eta}

\newcommand{\flow}{u}
\newcommand{\force}{F}

\newcommand{\energy}{U}
\newcommand{\potentialStrength}{\epsilon}
\newcommand{\inverseDebyeScreeningLength}{\kappa}

\newcommand{\diameter}{d}

\newcommand{\diffusion}{D_\mathrm{0}}

\newcommand{\brownianTime}{\tau_\mathrm{B}}
\newcommand{\boltzmannConstant}{k_\mathrm{B}}
\newcommand{\temperature}{T}
\newcommand{\numberOfParticles}{N}

\newcommand{\numberOfEnsemebleSystems}{N_\mathrm{ens}}

\newcommand{\azimuthalAngle}{\varphi}
\newcommand{\radius}{r}

\newcommand{\meanSquaredDisplacement}{\mathrm{MSD}}
\newcommand{\magneticField}{B}
\newcommand{\angularVelocity}{\omega}
\newcommand{\susceptibility}{\chi}
\newcommand{\volume}{V}
\newcommand{\magneticConstant}{\tilde{\mu}_\mathrm{0}}

\newcommand{\torque}{\bm{\mathcal{T}}}
\newcommand{\heat}{q}
\newcommand{\work}{w}

\newcommand{\dipoleMoment}{\bm\mu}

\newcommand{\stiffness}{K}
\newcommand{\Diff}{\mathrm{d}}
\newcommand{\probability}{P}
\newcommand{\standardDeviation}{\sigma}

\newcommand{\stochasticMoment}{m}
\newcommand{\magneticFieldFrequency}{\omega_\mathrm{M}}

\newcommand{\unityMatrix}{\mat{I}}
\newcommand{\oseenTensor}{\mat{G}^\mathrm{O}}

\newcommand{\blakeTensor}{\mat{G}^\mathrm{B}}

\newcommand{\leviCivitaTensor}{\mat{\underline{\hat\epsilon}}}

\begin{document}



\title{Dynamical modes of sheared confined microscale matter}

\author{Sascha Gerloff}
\email{s.gerloff@tu-berlin.de}
\affiliation{%
 Institut f\"ur Theoretische Physik, Hardenbergstr. 36,\\ Technische Universit\"at Berlin, D-10623 Berlin, Germany
}%
\author{Antonio Ortiz-Ambriz}
\affiliation{
Departament de Física de la Matèria Condensada, Universitat de Barcelona, Barcelona 08028, Spain}
\affiliation{
Institut de Nanociència i Nanotecnologia (IN2UB), Universitat de Barcelona, Barcelona 08028, Spain}
\author{Pietro Tierno}
\affiliation{
Departament de Física de la Matèria Condensada, Universitat de Barcelona, Barcelona 08028, Spain}
\affiliation{
Institut de Nanociència i Nanotecnologia (IN2UB), Universitat de Barcelona, Barcelona 08028, Spain}
\affiliation{
Universitat de Barcelona Institute of Complex Systems (UBICS), Barcelona 08028, Spain}
\author{Sabine H. L. Klapp}
\email{klapp@physik.tu-berlin.de}
\affiliation{%
 Institut f\"ur Theoretische Physik, Hardenbergstr. 36,\\ Technische Universit\"at Berlin, D-10623 Berlin, Germany
}%

\date{\today}

\begin{abstract}
Based on (overdamped) Stokesian dynamics simulations and video microscopy experiments, we study the non equilibrium dynamics of a sheared colloidal cluster, which is confined to a two-dimensional disk.
The experimental system is composed of a mixture of paramagnetic and non magnetic polystyrene particles, which are held in the disk by time shared optical tweezers.
The paramagnetic particles are located at the center of the disk and are actuated by an external, rotating magnetic field that induces a magnetic torque.
We identify two different steady states by monitoring the mean angular velocities per ring.
The first one is characterized by rare slip events, where the inner rings momentarily depin from the outer ring, which is kept static by the set of optical traps.
For the second state, we find a bistability of the mean angular velocities, which can be understood from the analysis of the slip events in the particle trajectories.
We calculate the particle waiting- and jumping time distributions and estimate a time scale between slips, which is also reflected by a plateau in the mean squared azimuthal displacement.
The dynamical transition is further reflected by the components of the stress tensor, revealing a shear-thinning behavior as well as shear stress overshoots.
Finally, we briefly discuss the observed transition in the context of stochastic thermodynamics and how it may open future directions in this field.
\end{abstract}
%
\maketitle
\section{Introduction}\label{sec:introduction}
Understanding the response to shear of complex systems, such as emulsions, gels, polymeric solutions, foams, glasses and colloidal suspensions, is key for various applications \cite{Berthier2011, Bonn2017, Puertas2014}.
Placing such materials inside strong spatial confinement has severe impact on their response to external deformations, which is crucial for a multitude of applications such as thin-film lubrication \cite{Annunziata2016, Bhushan1995, Ma2016, Raviv2003}, microfluidic devices \cite{Atencia2005, Genovese2011} and colloidal machines at the microscale
\cite{DiLeonardo2010, Williams2016, Ortiz-Ambriz2018}, to name a few.
Further, the material response to shear is intimately connected to the non-equilibrium dynamics of the constituent elements, that have been the subject of recent research with non-Brownian particles \cite{Fornari2016, Yeo2010}, polymer-  \cite{Huang2014}, active bacteria- \cite{Wioland2016}, and colloidal suspensions in amorphous \cite{Cohen2004, Schall2010, Shrivastav2016}, fluid- \cite{Isa2009, Ramaswamy2017}, as well as crystalline states \cite{Gerloff2017, Lin2016, Mackay2014, Vezirov2013}.

Colloidal suspensions under external fields have proven to be a powerful test bed system, that is used to study the role of channel geometry \cite{Cohen2004,Genovese2011,Wilms2012} hydrodynamic interactions \cite{Mackay2014, Uspal2012}, frictional interparticle contact and lubrication \cite{Royer2016, Vinutha2016}, as well as plastic events \cite{Gerloff2016, McDermott2016, Horn2014}, to cite a few.
Key advantages of using colloidal particles are the possibilities to directly visualize the particle dynamics via video microscopy, and to tune the pair interactions using external fields \cite{Martinez-Pedrero2015, Straube2014}. Note that in dense systems, tracking the particle dynamics in the bulk may be challenging.
In this context, two-dimensional colloidal clusters represent a simple, yet non trivial, model system to visualize and investigate the rich many-body dynamics of strongly interacting microspheres under shear.

Recently, we used such a system to explore the rheological response for a large range of shear flow strengths \cite{Ortiz-Ambriz2018}.
The experimental system consists of an ensemble of microspheres, which are confined by optical forces in a two dimensional circular Couette shear cell.
The two confining "walls", consisting of colloidal particles, can be actuated independently of each other by using magnetic- and optical forces.
These forces give rise to a hydrodynamic shear flow and induce complex non-equilibrium behavior such as shear-thinning as well as local shear-thickening.

In the present study, we focus on much smaller strengths of the shear flow. We aim at analyzing the non-equilibrium dynamics related to the initial breaking of the equilibrium structure and the onset of net particle transport in detail.
We find that this onset of motion is characterized by two different steady states.
Importantly, we investigate not only the net particle transport inside these steady states but also their fluctuations both in experiments and simulations, which reveal a bistability for a large range of shear flow strengths.
We find very good agreement between the experiments and the numerical simulations by comparing the distribution of angular velocities per ring.
Further, we analyze the different dynamical modes which emerge upon shear, and observe a series of locking and slip events.
These slip events are reminiscent of the avalanche-like dynamics generally observed in several amorphous systems across different length scales \cite{Denisov2016, Papanikolaou2016}, from earthquakes \cite{Brinkman2016}, to strongly correlated systems \cite{Zhou2015}.
One aspect of particular interest is the waiting time between two slips as well as its duration.
These two fluctuating quantities characterize a typical time scale for the plastic events, that is reflected by the mean squared displacements (MSD) of the particles as well as the shear stress relaxations.
In particular, we find a characteristic plateau of the MSD as well as a shear stress overshoot, that is commonly observed for sheared glasses \cite{Zausch2008}.

Finally, we briefly discuss the consequences of our results for two important stochastic thermodynamics quantities, i.e. the work and heat, that describe the energy supplied from an external source, i.e. the magnetic field, as well as the energy dissipated into the bath, respectively.
Interestingly, the non-equilibrium transition between the two steady states is most clearly reflected by the heat distributions, displaying a marked behavior with respect to its mean and the strength of its fluctuations.
This is somewhat different to our findings for a planar slit pore system \cite{Gerloff2018}, where we have found the opposite, namely that a most marked response for the stochastic work distributions.\\

The paper is organized as follows.
In \sref{experimental system}, we describe the experimental setup that we model using Stokesian dynamics simulations, whose details we discuss in \sref{numerical calculations}.
We then continue to discuss the azimuthal dynamics per ring in \sref{mean azimuthal dynamics} as well as the corresponding microscopic dynamics in \sref{microscopic motion}.
The rheological response inside the different steady states is then characterized in \sref{stress tensor}.
Finally, we briefly discuss some results for the stochastic energetics in \sref{stochastic energetics} and provide some general conclusions.
\section{Experimental System}\label{sec:experimental system}
We assemble clusters of microspheres, as shown in \fref{nanoclutch setup}(a), by trapping polystyrene particles with time averaged optical tweezers.
The colloidal suspension is prepared by first mixing ultrapure water with a small amount of TMAH ($\sim7\rm{\mu M}$) to counteract the absorption of CO$_2$. After that, we add $1$ml of this solution to disperse $5\rm{\mu l}$ of stock solution of superparamagnetic particles ($4.5 \rm{\mu{}m}$ in diameter, M-450 Epoxy Dynabeads) and $50 \rm{\mu l}$ of carboxylate modified latex particles, ($4\rm{\mu{}m}$ in diameter, CML Molecular Probes).
The colloidal suspension is then sandwiched between two cover glasses which are sealed with parafilm and silicon vacuum grease.

Our optical tweezers setup consists of a $1064$nm laser (ManLight ML5-CW-P/TKS-OTS, operated at $3$W) deflected by an Acousto Optic Device (AOD, AA Optoelectronics DTSXY-400-1064, driven by a RF generator DDSPA2X-D431b-34 and two NI card cDAQ NI-9403) and focused from above by a microscope objective (Nikon $40\times$ CFI APO). The sample is observed from below through a second objective (Nikon  $40\times$Plan Fluor) which is projected to a CMOS camera (Ximea MQ003MG-CM).
The AOD moves the trap to a new position every $0.5$ms. This speed is such that, for a typical ring of $21$ particles, each trap is visited every $10$ms. This time scale is much smaller than the characteristic Brownian time of the particles ($\tau\sim40$s).
Thus, the potential generated by this \emph{time shared} optical trap can be considered quasi static and effectively acts simultaneously as $21$ individual traps for each particle, respectively.
We use a custom built LabVIEW program to observe and to manipulate the particles through a graphical interface, and to assemble the cluster one by one.
The LabVIEW code can be accessed via github \cite{githubref}.

The sample is placed inside a set of five coils which allow us to apply a magnetic field in any direction. The coils are driven by a NI card (cDAQ NI-9269), and the signal is amplified with three power operational amplifiers (KEPCO BOP 20-10). The magnetic field is also controlled by a LabVIEW program, which allows us to automate the data acquisition.
Particle positions are extracted using the trackpy \cite{Trackpy} implementation of the Crocker-Grier algorithm \cite{Crocker1996e}.
\section{Numerical Calculations}\label{sec:numerical calculations}
\subsection{Model}\label{sec:model}
\begin{figure}
  \centering\includegraphics[width=0.95\linewidth]{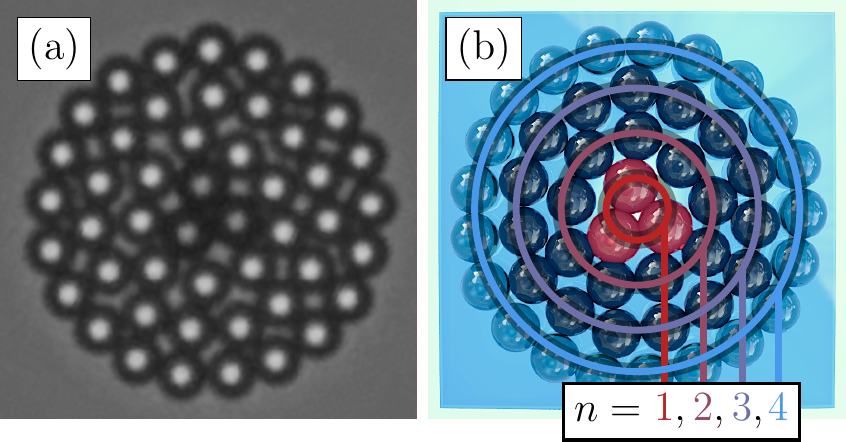}
  \caption{(a) Experimental and (b) schematic image of the considered system, consisting of a two dimensional cluster of confined microspheres. For clarity,we have marked in (b) the four rings $n=1,2,3,4$ with colored lines. The outer ring of particles ($n=4$) is trapped by optical tweezers, while the three paramagnetic particles, forming an inner triangle ($n=1$) at the center, are actuated by a rotating magnetic field.}
  \figLabel{nanoclutch setup}
\end{figure}
As shown in \fref{nanoclutch setup}, we consider a suspension containing a two-dimensional disk-like colloidal cluster.
It consists of $45$ polystyrene and $3$ paramagnetic particles, whose diameter are $\diameter = 4\,\mu\mathrm{m}$ and $\diameter_\mathrm{M} = 1.125\diameter$, respectively.
Similar to the experimental setup, the colloids inside the cluster are confined by an outer ring of polystyrene particles, and each particle is subjected  to a harmonic trap potential, which is given by
\begin{equation} \eqLabel{harmonic trap potential}
  \energy_\mathrm{R}\lr{\bv{\radius}_i, t} = \frac{\stiffness}{2} \abs{ \bv{\radius}_i - \bv{\radius}_{i,0}}^2\text{,}
\end{equation}
where $\bv\radius_i$ is the position of particle $i\in\cLR{1,\ldots,\numberOfParticles_4}$ of the outer ring, consisting of $\numberOfParticles_4 = 21$ particles, $\stiffness = 6000 \boltzmannConstant\temperature/\diameter^2$ is the stiffness of the harmonic trap and $\bv{\radius}_{i,0}$ its center.
In this study, we focus on the special case where the position of the harmonic traps is fixed at an equiangular distance on a ring $\displaystyle \bv{\radius}_{i,0} = R_\mathrm{out} \LR{ \cos\lr{ \Phi_{i}} \bv{e}_x + \sin\lr{ \Phi_{i} } \bv{e}_y }$, where $R_\mathrm{out} = 3.52\diameter$ is the radius of the outer ring and $\Phi_{i}=2\pi\, i / \numberOfParticles_4 $ is the angular position of particle $i$.

The inner paramagnetic particles are driven by a rotating magnetic field $\displaystyle \bv\magneticField\lr{t} = \magneticField_0 \LR{ \cos\lr{\magneticFieldFrequency}\unitVector{x} + \sin\lr{\magneticFieldFrequency}\unitVector{y} }$, where $\magneticField_0$ is the strength of the magnetic field,
$\magneticFieldFrequency = 125.7 \mathrm{rad}\, \mathrm{s}^{-1}$ is the angular frequency employed in the experiments, and $\unitVector{x}$ as well as $\unitVector{y}$ are the unit vectors in $x$ and $y$ direction.
This magnetic field induces a finite internal relaxation time of the particle magnetization \cite{Martinez-Pedrero2015}, which can be modeled
via a simple relaxation equation $\displaystyle \partial\dipoleMoment_i\lr{t}/\partial t = \tau_\mathrm{rel}^{-1}\LR{ \dipoleMoment_i - \volume\susceptibility\bv\magneticField\lr{t} }$, where $\dipoleMoment_i$ is the dipole moment, $\volume = \pi\diameter^3 / 6$ the volume, and $\susceptibility = 1.4$ the susceptibility of the paramagnetic particle $i$, whereas
$\tau_\mathrm{rel} = 0.00015 \brownianTime$ is the relaxation time scale for the induced magnetic dipole and $\brownianTime = \diameter^2/\diffusion = 40\,\mathrm{s}$ is the Brownian time of a particle of diameter $\diameter$ that is defined by $\diffusion$ the diffusion constant.
As a result, the paramagnetic particles are subject to a net  magnetic torque $\torque_i$, defined via
\begin{equation}\eqLabel{magnetic torque}
    \torque_i = \frac{\magneticFieldFrequency}{2\pi}\int_0^{\frac{\magneticFieldFrequency}{2\pi}}\dipoleMoment_i\lr{t}\times\bv\magneticField\lr{t} \;\Diff t = \frac{ \volume\susceptibility\magneticField_0^2\magneticFieldFrequency\tau_\mathrm{rel} }{1 +\magneticFieldFrequency^2\tau_\mathrm{rel}^2} \,\unitVector{z}\text{,}
\end{equation}
which is proportional to $\magneticField_0^2$, the square magnetic field strength.

The steric particle-particle interaction between the polystyrene as well as paramagnetic particles is modeled via a generic \emph{Yukawa}-like potential, given by
\begin{equation}\label{eq:yukawa potential}
  \energy_\mathrm{Y}\lr{\radius_{ij}} = \potentialStrength_\mathrm{Y}\diameter\frac{\exp\lr{-\inverseDebyeScreeningLength\, \radius_{ij}}}{\radius_{ij}}\text{,}
\end{equation}
where $\potentialStrength_\mathrm{Y} = 1.6 \inverseDebyeScreeningLength \diameter \exp\lr{ \inverseDebyeScreeningLength\diameter } \boltzmannConstant\temperature \diameter$ is the strength of the particle interactions, $\inverseDebyeScreeningLength = 40 \diameter^{-1}$ is the inverse Debye screening length, $\diameter$ is the mean diameter $\diameter=\lr{\diameter_i+\diameter_j}/2$ of the interacting particles, $\radius_{ij}$ is the distance between the interacting particles.

The interaction between the induced dipole moments is modeled via a time-averaged dipole-dipole interaction exerted between the rotating paramagnetic particles, given by
\begin{equation}\eqLabel{mean dipole dipole potential}
    \energy_\mathrm{DD}\lr{\radius} = -\frac{\magneticConstant \susceptibility^2 \volume^2 \magneticField_{0,\mathrm{DD}}^2}{8\pi \radius^3}\text{,}
\end{equation}
where $\magneticConstant$ is the magnetic constant and $\displaystyle \magneticField_{0,\mathrm{DD}} =\; 0.117\,\LR{\mathrm{mT}}$ is a constant magnetic field strength.
The latter is set such that the radius of the inner ring consisting of paramagnetic particles is approximately constant and equal to that measured in experiments $R_\mathrm{in} = 0.625\diameter$.
\subsection{Simulation Details}\label{sec:simulation details}
We perform (overdamped) Stokesian dynamics simulations to investigate the non-equilibrium dynamics of the colloidal particles actuated by the magnetic torque.
The equation of motions read
\begin{widetext}
  \begin{equation}
    \eqLabel{stokesian dynamics nanoclutch}
    \frac{\partial \bv\radius_i \lr{ t } }{\partial t} =  \sum_{j=1}^{\numberOfParticles}\cLR{ \mat{M}_{ij}^\mathrm{TT}\cdot\LR{ \sum_{j\neq k} \bv\force_{jk}\lr{\radius_{jk}} + \bv\force_\mathrm{R}\lr{\bv\radius_j,t}  } + \mat{M}_{ij}^\mathrm{TR}\cdot\bv\torque_j}+ \frac{\partial\bv{W}_i}{\partial t}\text{,}
  \end{equation}
\end{widetext}
where $\bv\radius_i$ is the position of particle $i$, $\bv\force_{jk}$ is the interaction force stemming from \eref{yukawa potential} and (\ref{eq:mean dipole dipole potential}) and $\radius_{jk}$ is the distance between particle $j$ and $k$,
$\bv\force_\mathrm{R}$ is the force of the traps acting on the outer ring and resulting from \eref{harmonic trap potential}, $\bv\torque_j$ is the magnetic torque acting on the paramagnetic particles.
In addition, the colloids are subject to random displacements $\partial \bv{W}$ with zero mean and variance $2\diffusion \partial t$.

In our framework, the hydrodynamic interactions between the particles are accounted for via the translation-translation $\displaystyle \mat{M}_{ij}^\mathrm{TT}$ as well as the translation-rotation mobility tensors $\displaystyle \mat{M}_{ij}^\mathrm{TR}$, see \eref{translation translation self mobility} and \eref{translation rotation self mobility} given in Appendix A.
In particular, we use expressions that include the finite extent of the colloidal particles on the Rotne-Prager level as well as the presence of the plane boundary represented by the bottom of the specimen, see Appendix A for details.
Note that, compared to Ref. \cite{Ortiz-Ambriz2018}, we here employ new, refined expressions for the $\displaystyle \mat{M}_{ij}^\mathrm{TT}$ in order to treat the bidispersity of the considered colloidal suspension accurately.
As a consequence, we have identified a new set of parameters for the particle interactions ($\potentialStrength_\mathrm{Y}$ and $\inverseDebyeScreeningLength$) via a parameter scan that aims to match the mean dynamics from simulations and experiments.
To this end, we consider the limiting case of a vanishing magnetic field $\magneticField_0 = 0$ and rotate the outer ring with constant angular velocity $\Phi_{i} = \angularVelocity_\mathrm{R}\, t + 2\pi i /\numberOfParticles_4$, a situation which was discussed in Ref. \cite{Ortiz-Ambriz2018} as well as \cite{Williams2016} for a monodisperse cluster.
For this limiting case, we compute and compare the mean angular velocities per ring as a benchmark to identify an appropriate set of parameters.
\section{Azimuthal Dynamics}\label{sec:mean azimuthal dynamics}
\begin{figure}
  \centering\includegraphics[width=0.95\linewidth]{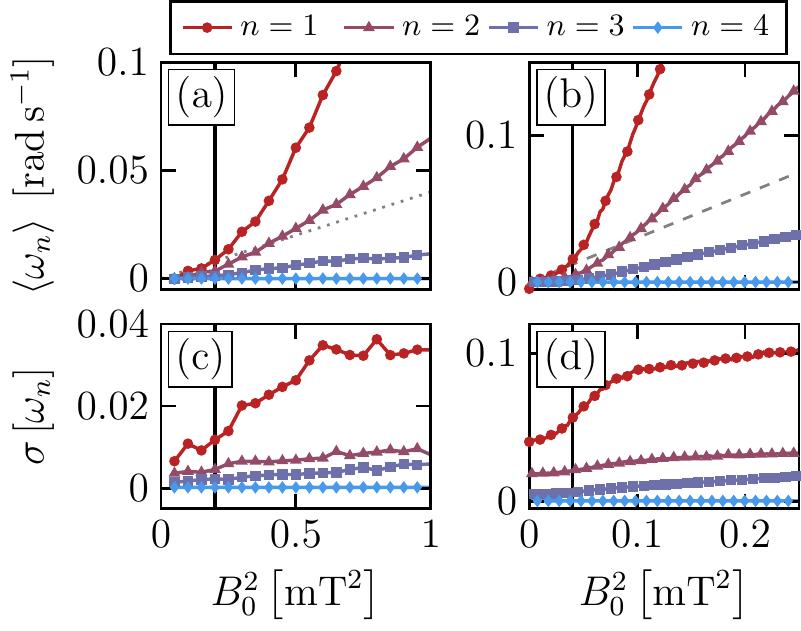}
  \caption{(a) Experiment and (b) simulation results for the ensemble averaged mean angular velocity per ring $n=1,\ldots,4$ as a function of the magnetic torque $ \propto \magneticField_0^2$. The gray dotted line in (a) indicates the initial slope of the angular velocity of the paramagnetic particles, $0.04\,\magneticField_0^2$, in experiments, and the dashed line in (b) indicates the corresponding slope $0.3\,\magneticField_0^2$ in simulations. The standard deviations of the angular velocities in (a) and (b) are plotted in (c) and (d), respectively.}
  \figLabel{mean angular velocity}
\end{figure}
In our previous study \cite{Ortiz-Ambriz2018}, we have focused on the particle dynamics as well as the corresponding rheological response at rather strong magnetic torques $\magneticField_0^2$.
We now concentrate on the dynamics that are observed for much smaller $\magneticField_0^2$.
Consequently, we limit ourselves to a much smaller range of field strengths $\magneticField_0 < 1 [\mathrm{mT}]$.

To characterize the steady state dynamics, we calculate the mean angular velocity per ring
\begin{equation}\label{eq:mean angular velocity}
  \angularVelocity_n = \qd{ \frac{1}{\numberOfParticles_n}\sum_i^{\numberOfParticles_n} \frac{ \varphi_i \lr{t+\Diff t} - \varphi_i\lr{t} }{2\pi\,\Diff t} }\text{,}
\end{equation}
where $\qd{\,\cdot\,}$ is a time average, $\numberOfParticles_n$ is the number of particles inside the $n\,$th ring ($\numberOfParticles_1=3$, $\numberOfParticles_2 = 9$, $\numberOfParticles_3 = 15$, $\numberOfParticles_4 = 21$), and $\varphi_i$ is the azimuthal angle defined via the relationship:
\begin{equation}
  \bv\radius_i\lr{t} = \radius\LR{\cos\lr{\varphi_i} \unitVector{x} + \sin\lr{\varphi_i}\unitVector{y}}\text{.}
\end{equation}
\subsection{Experiments}\label{sec:experiments}
In the experiments, the mean angular velocity is computed from $N_\mathrm{loops} = 60$ subsequent sweeps.
Starting from equilibrium, each sweep proceeds by slowly increasing the magnetic field, and thus the torque,  in discrete steps $\Delta\magneticField_0^2 = 0.05\,\mathrm{mT}^2$ up to a maximum of $\magneticField_0^2 = 1\,\mathrm{mT}^2$, followed by another sweep where the magnetic torque is decreased at the same rate to equilibrium.
The total duration of each sweep is $ 5\, \brownianTime \approx 200\,\mathrm{s}$.
Finally, averaging over all realizations, we find the mean angular velocity per ring, which is plotted in \fref{mean angular velocity}(a).
Here, we average over both, forward and backward, sweeps as we find that the mean angular velocities from the two sweeps are approximately the same.
This indicates that, at each step, the system managed to relax to the steady state and the sweeps were performed sufficiently slowly.

Starting from equilibrium and applying the magnetic field to the paramagnetic particles, the mean angular velocity of the inner ring first increases as a linear function of the torque, i.e. $\angularVelocity_1 \propto \magneticField_0^2$.
The initial slope is emphasized by the gray dotted line in \fref{mean angular velocity}(a).
At a critical magnetic torque $\magneticField_{0,\mathrm{c}}^2 = 0.2\,\LR{\mathrm{mT}^2}$ the inner ring speeds up, yielding another linear increase with larger slope.
Henceforth, we refer to this behavior as a "depinning transition" between two states with strongly different dependency of $\angularVelocity_n$ on the magnetic torque.
While overall slower, the second ($\angularVelocity_2$) and the third ring ($\angularVelocity_3$) both show the same behavior, including the depinning transition at the same critical magnetic torque.
Obviously, the outer ring remains static since the composing particles are trapped by the laser trap, for all $\magneticField_0^2$.
Note that a linear relation between the angular velocity and the magnetic torque is already found for a free rotating triplet of paramagnetic particles, which forms the inner ring, as discussed in the supplementary material of Ref. \cite{Ortiz-Ambriz2018}.
However, the actual slope and magnitude of the resulting angular velocities per ring strongly depend on the interactions between them.

Importantly, in contrast to Ref. \cite{Ortiz-Ambriz2018}, we find that for $\magneticField_0^2 < \magneticField_{\mathrm{c},0}$ we do not observe a fully locked state, i.e. a state where $\angularVelocity_n$ vanishes completely and the particles remain static on average. The reason is that in our previous work
the dynamics at small $\magneticField_0^2$ were not sufficiently resolved to distinguish between a static state and the very small mean angular velocity, as shown in \fref{mean angular velocity}(a) and (b).
However, for $\magneticField_0^2 < \magneticField_{0,\mathrm{c}}^2$, the inner rings do \emph{not} perform a regular rotation either.
Instead, the slow mean angular motion results from a series of slip events, where the inner rings "depin" from the static outer ring for a brief moment before locking again.
This behavior is reminiscent of the behavior near the depinning transition in incommensurate driven monolayers at finite temperature \cite{Hasnain2013}.
For the latter, one also observes a small net particle flux for subcritical driving forces, due to rare particle jumps that are induced by the thermal noise.
The slip events in the present system are discussed in more detail in \sref{microscopic motion}.

The depinning transition at $\magneticField_{\mathrm{c},0}^2$ is also reflected by the standard deviation of the mean angular velocities
\begin{equation}\eqLabel{standard deviation}
\standardDeviation\LR{ \angularVelocity_n } = \sqrt{\qd{\lr{\angularVelocity_n-\qd{\angularVelocity_n}}^2}}\,\mathrm{,}
\end{equation}
as shown in \fref{mean angular velocity}(c), which displays a marked increase at the transition.
\subsection{Simulations}\label{sec:simulations}
In simulations, we mimic the procedure employed in the experiments performing up to $N_\mathrm{ens} = 10000$ forward and backward sweeps with $\Delta \magneticField_0^2 = 0.0025\,\mathrm{mT}^2$ up to a maximum of $\magneticField_0^2 = 0.25\,\mathrm{mT}^2$ and a total duration of $10\,\brownianTime$ per sweep.
Similar to the experiments, we find a transition at $\magneticField_{0,\mathrm{c}}^2$ between two dynamical states, both characterized by a linear increase of the mean angular velocity.
In simulations the initial slope is indicated by a gray dashed line, see \fref{mean angular velocity}(b).
We find that the critical magnetic torque in simulations, $\magneticField_{0,\mathrm{c}}^2 = 0.04\,\mathrm{mT}^2$, is smaller than in experiments, and this difference in the applied magnetic field was also reported previously \cite{Ortiz-Ambriz2018}.
At the same time, for $\magneticField_0^2 > \magneticField_{0,\mathrm{c}}^2$, the mean angular velocity of the inner three rings is larger than that of the experiments.
We attribute these deviations to the limitations of our approximations for the hydrodynamic interactions as well as the fact that we neglect the surface friction between the rotating colloidal particles.

In addition to the mean values, the standard deviation of the mean angular velocity $\standardDeviation\LR{\angularVelocity_n}$, plotted in \fref{mean angular velocity}(d), is in good agreement with that of the experiments, showing a marked increase at the critical magnetic torque $\magneticField_{0,\mathrm{c}}^2$.
However, in simulations the mean angular velocity fluctuations are much stronger than in experiments.\\

\begin{figure}
  \centering\includegraphics[width=0.95\linewidth]{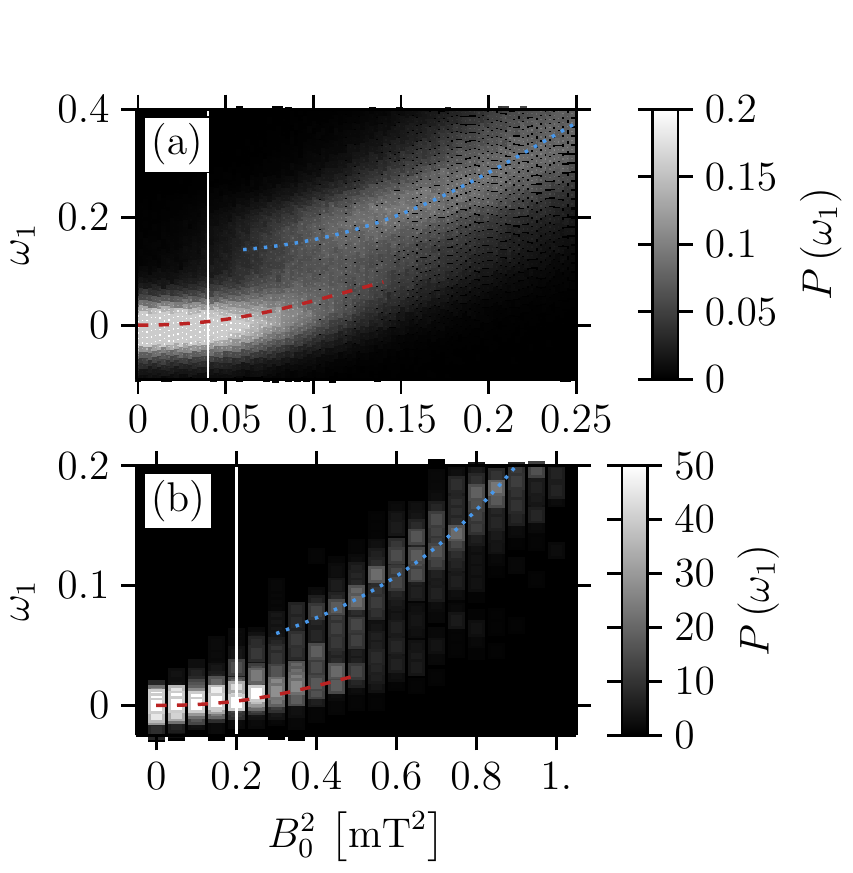}
  \caption{Evolution of the distribution of the mean angular velocity of the inner ring $\angularVelocity_1$ during the magnetic field sweeps from simulations (a) and experiments (b). The dashed red and dotted blue line indicate the evolution of the two peaks of $\probability\lr{\angularVelocity_1}$. The distributions correspond to the mean values and the standard deviations, which are plotted in \fref{mean angular velocity}(a-d).}
  \figLabel{mean angular velocity distribution}
\end{figure}
To further analyze the depinning transition at $\magneticField_{0,\mathrm{c}}^2$, we compute the distribution $\probability\lr{\angularVelocity_n}$ of the time-averaged angular velocities of the individual realizations \footnote{
  These distributions are different from that of the distribution of the instantaneous angular velocity per ring $\angularVelocity_n\lr{t}$ during each realizations. These tend to be much wider and featureless, i.e. Gaussian, due to strong spatial fluctuations, data not shown.
}, which are plotted in \fref{mean angular velocity distribution}(a) and (b) for the inner ring from simulations and experiments, respectively.
Again, we find that the width of the distributions, i.e. the standard deviation, displays a sudden increase at $\magneticField_{0,\mathrm{c}}^2$, as already shown in \fref{mean angular velocity}(c) and (d) for experiments and simulations, respectively.
The large width of the distributions reflects the fact that we find realizations, which seem to be momentarily locked with $\angularVelocity_n \approx 0$ even for $\magneticField_0^2 > \magneticField_{0,\mathrm{c}}^2$.
Overall, the distributions for the three inner rings $n=1,2,3$ are non-Gaussian, displaying positive skewness $\stochasticMoment_3\LR{ \angularVelocity_n } > 0$ as well as large kurtosis $\stochasticMoment_4\LR{\angularVelocity_n}>3$ for most $\magneticField_0^2$.
Here we have used the standard definition for the skewness
\begin{equation}\eqLabel{skewness}
  \stochasticMoment_3\LR{ x }=\frac{\qd{\lr{ x -\qd{x}}^3}}{\standardDeviation\LR{x}^3}
\end{equation}
as well as the kurtosis
\begin{equation}\eqLabel{kurtosis}
  \stochasticMoment_4\LR{x}=\frac{\qd{\lr{ x -\qd{x}}^4}}{\standardDeviation\LR{x}^4} \text{.}
\end{equation}
Thus, all stochastic moments clearly reflect the depinning transition at $\magneticField_{0,\mathrm{c}}^2$, where the skewness displays a maximum and the kurtosis a steep decrease (data not shown).

Further, we find that the distribution of the inner ring, $\probability\lr{\angularVelocity_1}$, display the most complex behavior.
In particular, we observe a range $\magneticField_0^2 = 0.06\,-\,0.14\,\LR{\mathrm{mT}^2}$ where $\probability\lr{\angularVelocity_1}$ becomes bimodal, see the red dashed and blue dotted line in \fref{mean angular velocity distribution}(a).
In contrast, $\probability\lr{\angularVelocity_2}$ and $\probability\lr{\angularVelocity_3}$ remain unimodal for all $\magneticField_0^2$, data not shown here.
In the bimodal regime some realizations are essentially "locked", i.e. $\angularVelocity_1\approx0$, whereas other realizations display a rotation with finite angular velocity, corresponding to a "running" state.
With increasing $\magneticField_0^2$ the number of realizations in an approximately locked state (red dashed) decreases and the number of realizations in the "running" state (blue dotted) increases continuously.
We note again that the distributions $\probability\lr{\angularVelocity_n}$ from the forward and the backward sweeps are approximately identical, indicating that the steady state for these magnetic torques is truly bistable.

In the experiments, we find very good agreement with the simulations results, as shown in \fref{mean angular velocity distribution}(b).
That is, we also find a range $\magneticField_0^2 = 0.3\,-\,0.6\,\LR{\mathrm{mT}^2}$ where $\probability\lr{\angularVelocity_1}$ becomes bimodal.
In this field range, we find realizations either in a locked or in a running state, where the number of realizations in a running state increases with increasing magnetic torque.
Thus, the agreement between simulation and experiments is not limited to the mean values but extends to the fluctuations of $\angularVelocity_n$.

To understand this bistability, recall the fact that each $\angularVelocity_n$ is computed from a sweep lasting $\Delta t = 0.1\brownianTime$ in simulations and $\Delta t = 0.25 \brownianTime$ in experiments for each magnetic torque.
Thus, $\probability\lr{\angularVelocity_1}$ shows that the rings of the single realization are either \emph{momentarily} ($t>\Delta t$) locked or running, corresponding to the two states respectively.
In fact, we do not find a single realization that remains in either the locked nor the running state for all times.
As a result, the bistability is intimately connected to the microscopic dynamics, which we discuss below.
\section{Microscopic (Angular) Dynamics}
\label{sec:microscopic motion}
The average azimuthal dynamics, that we have discussed in \sref{mean azimuthal dynamics}, is intimately related to the dynamics of the individual colloids.
In this section, we analyze the latter by means of the particle trajectories, the waiting- and jump time distributions, as well as the mean squared displacement.
\subsection{Trajectories}\label{sec:trajectories}
\begin{figure}
  \centering\includegraphics[width=0.95\linewidth]{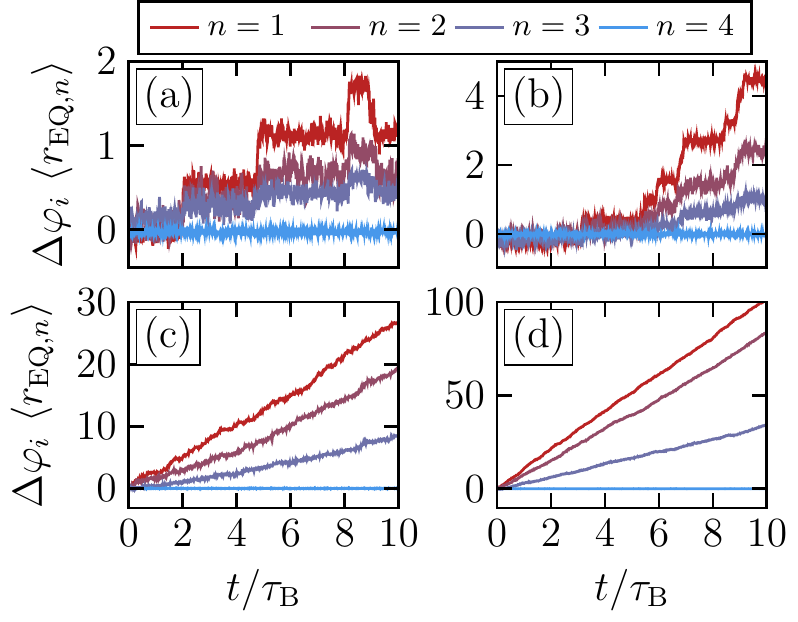}
  \caption{(a)-(d) Rescaled angular displacement trajectories for one arbitrary particle belonging to the $n=1,2,3,4$ ring for four different magnetic torques $\magneticField_0^2 = 0.0025, 0.04,0.1, 0.25 \;\LR{\mathrm{mT}^2}$, which are below, at, and above the critical value $\magneticField_{\mathrm{c},0}^2=0.04\,\LR{\mathrm{mT}^2}$. To improve clarity, we have multiplied the angular displacement for each particle by the mean radius of their ring at equilibrium ($\radius_{\mathrm{EQ},n}$). The trajectories therefore represent approximately the azimuthal displacement of the particle.}
  \figLabel{angular displacement trajectories}
\end{figure}
In \fref{angular displacement trajectories}(a)-(d), we have plotted simulation results for trajectories of the single particles.
At very small magnetic torques $\magneticField_0^2 = 0.0025 \LR{\mathrm{mT}^2}$, see \fref{angular displacement trajectories}(a), we find long periods where the particles are locked to their respective angular position (i.e., $\Delta \azimuthalAngle_i\lr{t} = \azimuthalAngle_i\lr{t} - \azimuthalAngle_i\lr{0} = \mathrm{const.}$), which are disrupted by sudden and very fast slips.
At the end of a slip the particle then resides again at an approximately constant $\Delta \azimuthalAngle_i\lr{t}$ for relatively long periods.
At the magnetic torques considered, these slips are likely to be triggered by thermal fluctuations, allowing also for "backward" slips where particles jump to smaller values of $\Delta \azimuthalAngle_i$.
These slips lead to a very small, but finite, mean angular  velocity, as shown in \fref{mean angular velocity}.

Increasing the magnetic torque, the frequency of the slips increases and the angular displacement during a single slip becomes larger.
This is clearly reflected in the trajectories at $\magneticField_0^2 = \magneticField_{\mathrm{c},0}^2=0.04\,\LR{\mathrm{mT}^2}$, see \fref{angular displacement trajectories}(b).
Here, we find again long periods, where the particles are locked, and a series of very fast slips.
However, the slips become more directed into positive angular direction $\Delta\azimuthalAngle_i > 0$ and the time between these slips becomes shorter.

For supercritical magnetic torques, e.g. $\magneticField_0^2 = 0.1 \,\LR{ \mathrm{mT}^2 }$, we find that the particle trajectories are mostly characterized by a continuous motion along azimuthal direction, see \fref{angular displacement trajectories}(c).
However, there are short periods where the inner rings lock and $\Delta \azimuthalAngle_i$ remains approximately constant.
While this transient locking is most pronounced for the inner ring ($n=1$), we observe similar behavior in the other rings $n=2,3$.
The transient locked periods reflect, on the trajectory level, the bistable region shown in \fref{mean angular velocity distribution}(a).
In particular, realizations inside these locked periods correspond to the low mobility states, i.e. red dashed line in \fref{mean angular velocity distribution}(a).

For large magnetic torques, i.e. $\magneticField_0^2 = 0.25 \,\LR{ \mathrm{mT}^2 }$, all the inner rings display a continuous motion in azimuthal direction, which is reflected by a continuous increase of $\Delta \azimuthalAngle_i$, as shown in \fref{angular displacement trajectories}(d).
Thus, the system has entered a running state, which
is consistent with our observations for the mean angular velocities of the inner rings \fref{mean angular velocity}(b).

\begin{figure}
    \centering\includegraphics[width=0.95\linewidth]{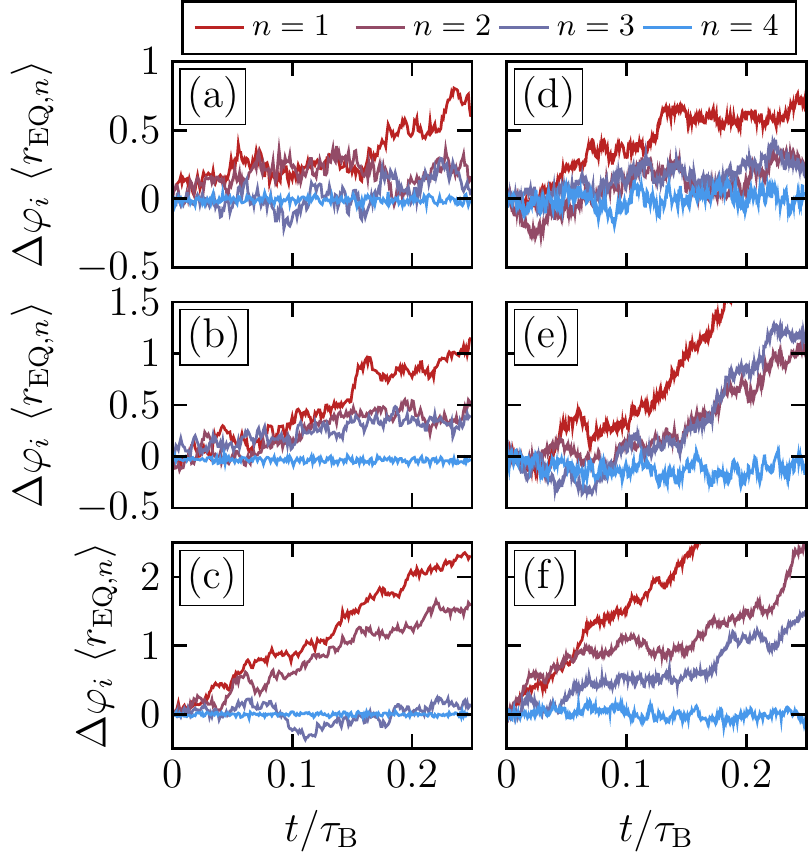}
  \caption{(a)-(c) Short time trajectories for one arbitrary particle of each ring for three different magnetic torques $\magneticField_0^2 = 0.04,0.1,0.25 \,\LR{\mathrm{mT}^2}$. (d)-(f) Trajectories from experiments for comparable magnetic torques $\magneticField_0^2 = 0.2, 0.4,1\,\LR{\mathrm{mT}^2}$. Thus, (a) and (d) correspond to the trajectories at $\magneticField_{0,\mathrm{c}}$.}
  \figLabel{angular displacement trajectories short}
\end{figure}
In \fref{angular displacement trajectories short}(a) and (d), we compare short trajectories from simulation and experiments, respectively. In experiments, the available particle trajectories are too short ($t=0.25\brownianTime$) to see the rare jumps for $\magneticField_0 < \magneticField_{0,\mathrm{c}}$,
However, for $\magneticField_0 > \magneticField_{0,\mathrm{c}}$ we find good agreement with particles showing short periods of transient locking as seen in simulations, see \fref{angular displacement trajectories short}(b), (c) and (e), (f).
\subsection{Waiting Time Distribution}\label{sec:waiting time distribution}
\begin{figure*}
  \centering\includegraphics[width=0.95\linewidth]{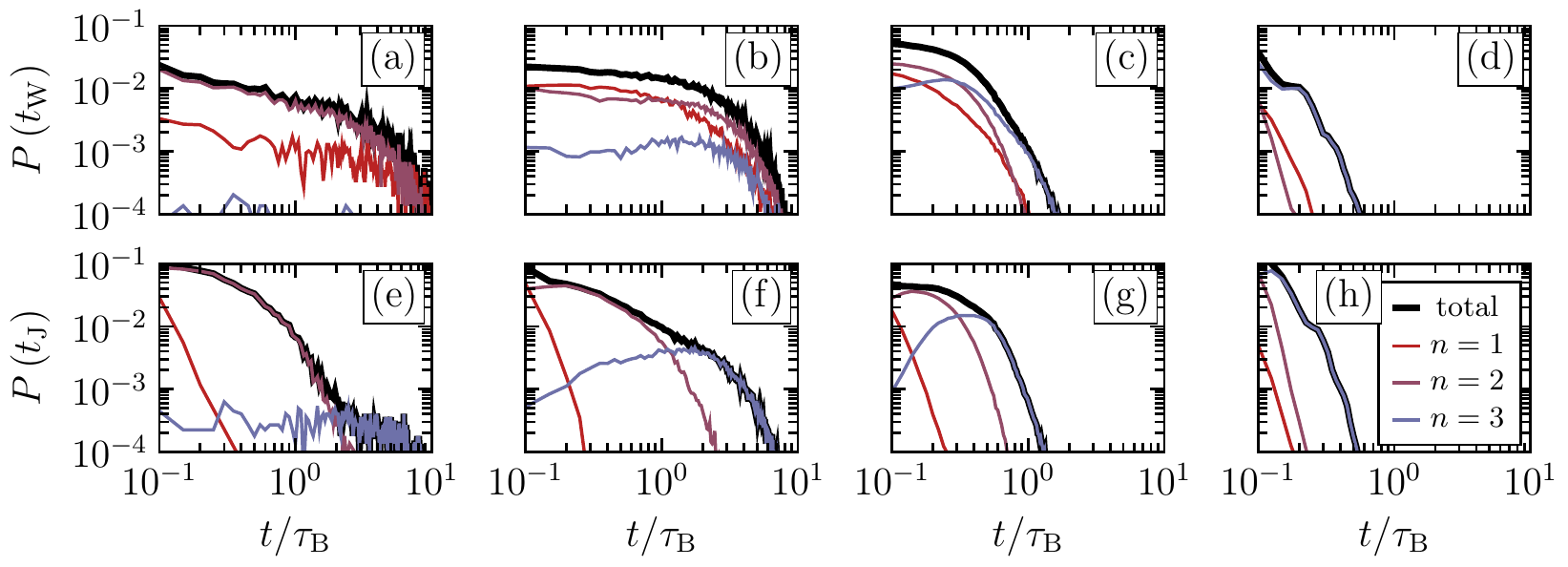}
  \caption{Numerical results for the waiting times (a-d) and jumping times (e-h) corresponding to the trajectories as shown in \fref{angular displacement trajectories}(a-d), for magnetic torques $\magneticField_0^2 = 0.0025,0.04, 0.1, 0.25\;\LR{\mathrm{mT}^2}$, respectively. The total distribution (black) is decomposed into that of the individual rings $n=1,2,3$. }
  \figLabel{waiting time distribution}
\end{figure*}
To quantify the frequency of the slip events, that we have discussed for \fref{angular displacement trajectories}(a-d), as well as their duration, we have computed the waiting- and jumping time distribution employing the definition from Ref. \cite{Gernert2014}.
We define, for each particle, a minimum angular position by the angle centered between the two neighboring particles of the outer adjacent ring.
A forward jump is initiated when a particle passes this minimum angular position along the positive direction.
The jump concludes after the jumping time $t_\mathrm{J}$ once the particle passes the minimum angular position of the next neighbors.
The time in between jumps is the waiting time $t_\mathrm{W}$.
We have plotted the corresponding waiting time- and jumping time distributions for $\magneticField_0^2 = 0.0025,0.04, 0.1, 0.25\;\LR{\mathrm{mT}^2}$ in \fref{waiting time distribution}(a-d) and (e-h), respectively.

Starting with the distribution $\probability\lr{t_\mathrm{W}}$, we find that $t_\mathrm{W}$ is approximately exponentially distributed for the first and second ring ($n=1,2$).
In particular, we find a high probability for short waiting times $t_\mathrm{W} < \brownianTime$, that are not apparent in the trajectories, see \fref{angular displacement trajectories}(a-b).
These short waiting times stem from particles jumping more than one interstice at once, which in our definition is interpreted as multiple subsequent jumps following each other, yielding very small waiting times between them.
In fact, especially for the rings with $n=1,2$, jumps over multiple minima seem to be very common.
With increasing magnetic torque, the waiting times become continuously shorter, which is consistent with the observations from the single particle trajectories, see \fref{angular displacement trajectories}.

Interestingly, the third ring ($n=3$) displays a different behavior, which we attribute to the fact that the corresponding particles jump only one interstice at a time.
For this ring, we observe in \fref{angular displacement trajectories}(b-d) distributions with pronounced maxima.
This maximum clearly indicates a characteristic time scale for the slips of the third ring, which again becomes shorter with increasing magnetic torque.
Comparing all the rings, we see that the typical waiting times in the range $\magneticField_0^2 \leq \magneticField_{0,\mathrm{c}}^2$ are approximately equal, indicating that the slip events are synchronized between the different rings.
In contrast, for larger magnetic torques, the waiting times become longer with increasing radial distance from the center.

For the jumping times, a dependency on the ring is observed for all magnetic torques.
Specifically, the first ring displays the smallest jumping times and the third ring the longest, see \fref{waiting time distribution}(c-d).
This difference is most prominent for $\magneticField_0^2 \leq \magneticField_{0,\mathrm{c}}^2$, where the typical duration of the jumps can vary by an order of magnitude.
Overall, most of the jumping time distributions display a maximum, which again yields a characteristic time scale of the microscopic dynamics.
\subsection{Mean Squared Angular Displacement}\label{sec:mean squared angular displacement}
\begin{figure}
  \centering\includegraphics[width=0.95\linewidth]{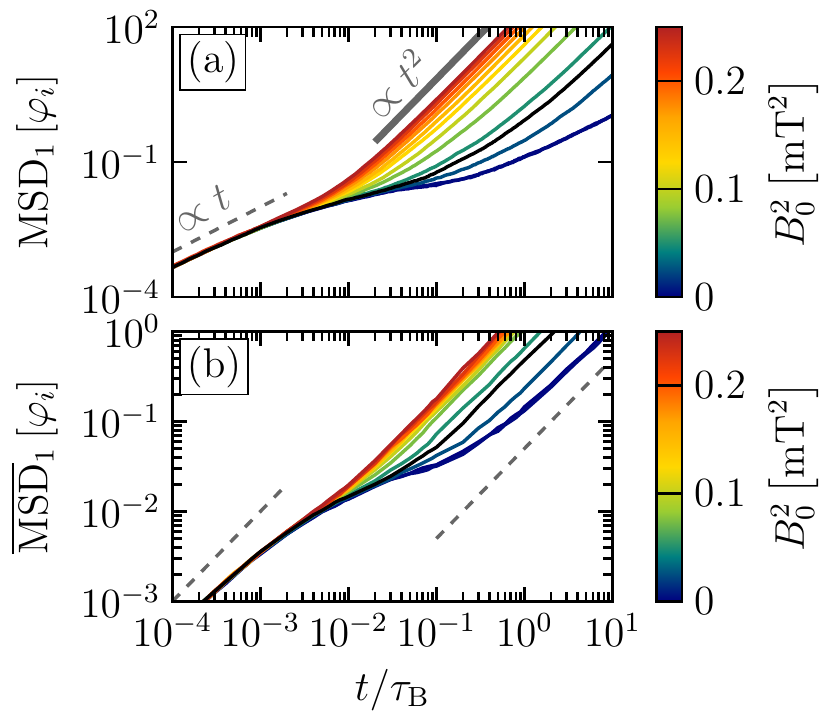}
  \caption{(a) Mean squared angular displacement of the particles inside the inner ring $n=1$ for various magnetic torques from simulations. (b) Mean squared displacement relative to the center of mass motion of the particles inside the inner ring $n=1$. The gray dashed line signifies a linear increase, whereas the gray solid line represents a quadratic increase.
  }
  \figLabel{mean squared angular displacement}
\end{figure}
A further measure of the microscopic motion, is the mean squared angular displacement (MSD), which is calculated from the individual trajectories
\begin{equation}\label{eq:mean squared displacement}
  \meanSquaredDisplacement_n\LR{ \azimuthalAngle_i } = \frac{1}{\numberOfParticles_n} \sum_i^{N_n} \qd{ \Delta\azimuthalAngle_i\lr{t}^2 }\text{,}
\end{equation}
where $\numberOfParticles_n$ is the number of particles of the $n\,$th ring, $\Delta\azimuthalAngle_i\lr{t}$ is the azimuthal angular displacement of particle $i$, $\qd{\,\cdot\,}$ is an ensemble average.
In addition, we consider the MSD, where the center of mass motion due to the external drive is subtracted
\begin{equation}\label{eq:mean squared displacement center of mass}
  \overline{\meanSquaredDisplacement}_n\LR{ \azimuthalAngle_i } = \frac{1}{\numberOfParticles_n} \sum_i^{N_n} \qd{ \LR{\Delta\azimuthalAngle_i\lr{t} - \qd{\Delta\azimuthalAngle_i\lr{t}} }^2 }\text{.}
\end{equation}
From now on, we focus on the mean squared angular displacement of the particles in the inner ring, $n=1$, but the behavior in the other rings is similar.
Results are plotted in \fref{mean squared angular displacement}(a) and (b).

In both cases, for very short times $t < 10^{-3}\brownianTime$, we find an initial linear increase of $\meanSquaredDisplacement\LR{\azimuthalAngle_i} \propto t$, corresponding to diffusive motion of the colloidal particles inside their respective "cages" formed by the surrounding colloidal particles.
At intermediate times $t \approx 0.01\brownianTime$, the mean squared angular displacement response becomes strongly dependent on $\magneticField_0^2$.
Considering first \fref{mean squared angular displacement}(a), we see that for small magnetic torques $\magneticField_0^2 < \magneticField_{\mathrm{c},0}^2$, the mean squared displacement displays a plateau up to $t\approx 0.1\,\brownianTime$.
For longer times, the MSD switches to ballistic motion, i.e. $\meanSquaredDisplacement\LR{\azimuthalAngle_i}\propto t^2$.
This corresponds to the directed angular motion that we have already discussed in \fref{mean angular velocity}(b).

Subtracting the center of mass motion, as plotted in \fref{mean squared angular displacement}(b), we see a similar transition of $\overline{\meanSquaredDisplacement}_1$.
However, $\overline{\meanSquaredDisplacement}_1$ transitions for long time again into a linear time dependency, corresponding to diffusive motion relative to the center of mass motion.
The corresponding diffusion constants increases continuously with the magnetic torque $\magneticField_0^2$.

Interestingly, the time range where $\meanSquaredDisplacement\LR{\azimuthalAngle_i}$ displays a plateau corresponds to the typical time in which the particles remain locked between the slips, as shown in \fref{angular displacement trajectories}(a) and (b).
With increasing magnetic torque, the width of the plateau decreases, corresponding to the decrease of the typical waiting times at the different $\magneticField_0^2$, shown in \fref{waiting time distribution}(a-d).
We note that such a plateau of the mean squared displacement is often observed in sheared colloidal glasses \cite{Zausch2008} and other strongly correlated driven fluids \cite{Emary2012, Laurati2012}.
For the former, one can relate the stress overshoots and the sub-diffusive domain with the breakage of the individual particle cages that are comprised of its neighboring particles \cite{Zausch2008}.
In fact, we do also observe stress overshoots as discussed below, thus we think that the same reasoning applies to our sheared colloidal system.
Here, the cages are composed of the particles from the neighboring rings as well as the direct neighbors inside the same ring.
\section{Stress Tensor}\label{sec:stress tensor}
To further characterize the observed dynamical behavior, we now discuss various mechanical properties.
In particular, following our previous study, we calculate the components of the configurational stress tensor in polar coordinates,
\begin{equation}\eqLabel{mean polar stress tensor}
  \stress_{nm} = \frac{1}{\volume} \qd{\sum_i^\numberOfParticles\sum_{j>i}^\numberOfParticles \LR{\bv\radius_{ij}\cdot\unitVector{n}\lr{\azimuthalAngle_{i}}}\,\LR{\bv\force_{ij}\cdot\unitVector{m}\lr{\azimuthalAngle_i}} }\text{,}
\end{equation}
where $n,m\in\cLR{\radius,\azimuthalAngle}$ are the polar coordinates, $\bv\radius_{ij}=\bv\radius_i-\bv\radius_j$ is the distance between two particles, $\bv\force_{ij}$ is the (interaction) force between particle $i$ and $j$, and $\unitVector{n}$, $\unitVector{m}$ are unit vectors in $n$- and $m$-direction.
\subsection{Shear Stress}
\begin{figure}
  \centering\includegraphics[width=0.95\linewidth]{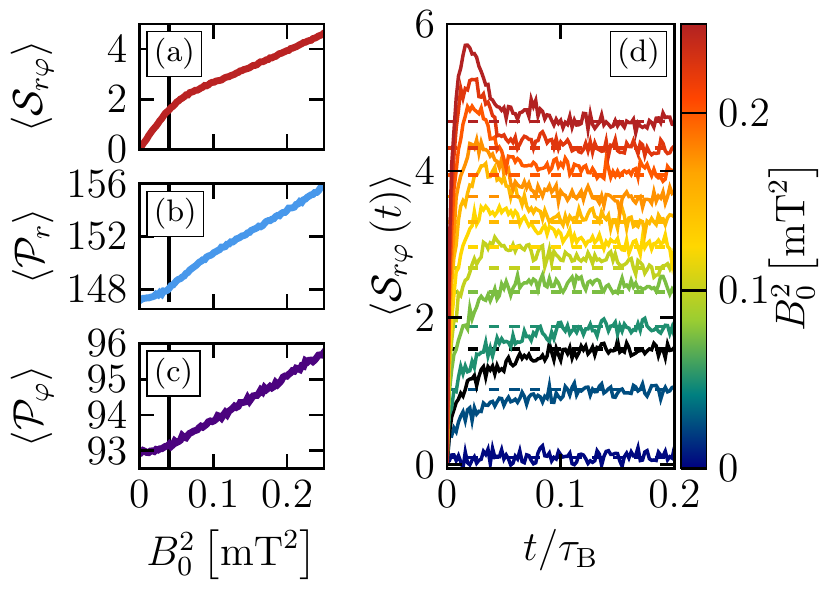}
  \caption{ (a) Mean shear stress $\stress_{\radius\azimuthalAngle}$, (b) mean radial pressure $\pressure_{\radius}$, (c) mean azimuthal pressure $\pressure_{\azimuthalAngle}$ as a function of the magnetic torque $\magneticField_0^2$. The black line indicates the critical magnetic torque at $\magneticField_{0,\mathrm{c}} = 0.04 \,\LR{\mathrm{mT}^2}$.
  (d) Shear stress relaxation curves $\stress_{\radius\azimuthalAngle}\lr{t}$ for various magnetic torques as indicated by the color. The black line corresponds to the stress relaxation at $\magneticField_{0,\mathrm{c}}$. The dashed lines correspond to the steady state shear stress plotted in (a).}
  \figLabel{shear stress}
\end{figure}
We start by considering the shear stress, $\stress_{\radius\azimuthalAngle}$, which is plotted in \fref{shear stress}(a).
Starting from equilibrium, the shear stress increases approximately linearly as a function of the magnetic torque up to $\magneticField_{0,\mathrm{c}}^2$, reflecting Newtonian behavior.
For supercritical magnetic torques $\magneticField_0^2 > \magneticField_{0,\mathrm{c}}^2$, the shear stress then crosses over to another linear increase with \emph{smaller} slope.
Note that the slope of the shear stress can be identified as the shear viscosity of the system \cite{Todd1995}; thus its decrease reflects a \emph{shear-thinning} behavior.

Similar to the fluctuations of the angular velocity, shown in \fref{mean angular velocity}(c-d), the shear stress displays fluctuations whose magnitude clearly reflect the depinning transition at $\magneticField_{0,\mathrm{c}}^2$.
In particular, we find a strong increase of the standard deviation as well as an increase of the skewness, data not shown here.
This effect implies that the stress fluctuations are increasingly biased towards large values, which can be attributed to the increasing number of slip events, during which large stresses are exerted.\\

We have also investigated the time-dependence of the shear stress, see \fref{shear stress}(d).
For various values of $\magneticField_0^2$ considered, we start in equilibrium and switch on the magnetic field at $t=0$ for up to $\numberOfEnsemebleSystems = 10000$ realizations.
The data reveal, first, that the relaxation towards the steady state values occurs in rather short times $t < 0.2\,\brownianTime$.
This is an indirect confirmation that the angular velocity sweeps described in \sref{simulations} are performed sufficiently slowly such that the internal stresses can relax.
Analyzing further the curves $\stress_{\radius\azimuthalAngle}\lr{t}$ in \fref{shear stress}(d), we find a monotonous increase of the shear stress within the range of $\magneticField_0^2 < \magneticField_{0,\mathrm{c}}^2$, that is the steady state shear stress is approached from below.
In contrast, for $\magneticField_0^2 > \magneticField_{0,\mathrm{c}}^2$, we find a non-monotonic behavior characterized by a pronounced stress "overshoot".
This overshoot dynamics is characteristic for glassy (and other strongly correlated) systems.
It is related to the breaking of the particles cages, consisting of neighboring particles \cite{Zausch2008}, and is connected to a plateau region in the mean squared displacement, see \sref{mean squared angular displacement}.
\subsection{Pressure}
\begin{figure}
  \centering\includegraphics[width=0.95\linewidth]{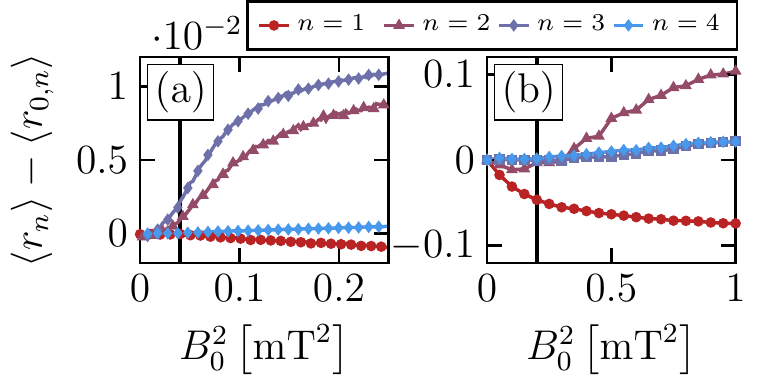}
  \caption{(a) Mean radius relative to the equilibrium radius for all rings ($n=1,\ldots,4$) from simulations and (b) experiments.}
  \figLabel{radial displacement}
\end{figure}
We now turn to the radial pressure, $\pressure_\radius= -\stress_{\radius\radius}$, and the azimuthal pressure, $\pressure_\azimuthalAngle = -\stress_{\azimuthalAngle\azimuthalAngle}$, which correspond to the diagonal components of the stress tensor and are plotted in \fref{shear stress}(b-c), respectively.
In general, both the radial- as well as the azimuthal pressure show similar behavior.
We find an approximately constant value for subcritical magnetic torques $\magneticField_0^2 < \magneticField_{0,\mathrm{c}}^2$, which crosses over to a continuous increase with torque for $\magneticField_0^2 > \magneticField_{0,\mathrm{c}}^2$.
Overall, the radial pressure is larger and grows faster than the azimuthal pressure, consistent with our results from previous studies \cite{Ortiz-Ambriz2018}.

The resulting increase of the radial pressure is accompanied by radial deformations of all rings from its equilibrium radius, which we have plotted in \fref{radial displacement}(a) and (b) for simulations and experiments, respectively.
In general, we observe a strong expansion of the system at the critical magnetic torque, where the particles start to slide past each other.
This leads to particles stacking up in radial direction, pushing the outer ring outwards against the (soft) harmonic traps, yielding eventually an increase of radial pressure.
Interestingly, in contrast to the three outer rings, the inner ring ($n=1$) displays a \emph{compression} for $\magneticField_0^2 > \magneticField_{0,\mathrm{c}}^2$.
These effects are seen in both, simulations and experiments.
However, the compression of the inner ring is much more pronounced in experiments, as shown in \fref{radial displacement}(b).
We attribute this to the fact that the strength of the induced dipole-dipole interaction is proportional to the magnetic torque, see \eref{mean dipole dipole potential}.
This results in a stronger attraction for larger magnetic torques.
As a result, the inner ring in the experimental system compresses already for $\magneticField_0^2 < \magneticField_{0,\mathrm{c}}^2$, which also leads to a small initial compression of the second ring.
In simulations, we keep the strength of the mean dipole-dipole interactions \eref{mean dipole dipole potential} constant to prevent a pronounced compression, which leads to a significant speedup of the inner ring that is not observed in experiments \cite{Ortiz-Ambriz2018}.
The observed compression of the inner ring $n=1$ is due to the inherent softness of the particle interactions: The outer particles push the inner particles inward.
Irrespective of these subtle differences between experiment and model system, the overall agreement is quite satisfactory.
\section{Thermodynamical Consequences}\label{sec:stochastic energetics}
In this last section we briefly discuss some aspects of the observed dynamical behavior from the perspective of stochastic thermodynamics.
Here, we focus on the stochastic energetics, i.e. the work and heat.
To this end, we employ generalized expressions for the stochastic work- and heat rate given in Ref. \cite{Speck2008}
\begin{align}
  \eqLabel{work rate}
  \dot\work\lr{t} &= \sum_{i=1}^\numberOfParticles \left\{ \left.\partialD{ \energy\lr{ \cLR{\bv\radius },t }}{t} \right|_{\bv\radius_i} -  \bv\flow\lr{ \bv\radius_i, t }\cdot \bv\force_i\lr{ \cLR{\bv\radius }, t } \right. \nonumber\\
&\left. \qquad\qquad +\; \bv\force_\mathrm{ext}\lr{\bv\radius_i}\cdot\LR{\partialD{\bv\radius_i}{t} - \bv\flow\lr{\bv\radius_i,t} }  \right\} \text{,} \\
  \eqLabel{heat rate}
    \dot\heat\lr{t} &=\sum_{i=1}^\numberOfParticles \LR{ \bv\force_\mathrm{ext}\lr{\bv\radius_i} + \bv\force_i\lr{ \cLR{\bv\radius}, t } } \cdot \LR{ \partialD{\bv\radius_i}{t} - \bv\flow\lr{\bv\radius_i,t} }\text{,}
\end{align}
where $\energy\lr{ \cLR{\bv\radius },t }$ is the total potential energy,
$\bv\flow\lr{ \bv\radius_i, t }$ is the external flow,
$\displaystyle \bv\force_i\lr{ \cLR{\bv\radius }, t } = \nabla_{\bv\radius_i}\energy\lr{ \cLR{\bv\radius },t }$ is the conservative force stemming from $\energy\lr{ \cLR{\bv\radius },t }$,
 and $\bv\force_\mathrm{ext}$ is an external force acting on the colloidal particles.
Note that, as already stated in Ref. \cite{Speck2008}, the hydrodynamic interaction enter only implicitly via the motion of the particles $ \partial\bv\radius_i / \partial t $.

For the sheared colloidal system with a static outer ring, $\partial \energy /\partial t = 0$ and $\bv\force_\mathrm{ext}=0$, we identify only one possible source of work.
This is the work done by the rotating magnetic field, which drives the rotation of the paramagnetic particles.
Within our model, we do not account for the rotational degrees of freedom of the colloidal particles explicitly.
Rather, the rotating magnetic field enters through the mean solvent flow exerted by the paramagnetic particles.
Therefore, we set the external flow in \eref{work rate} and \eref{heat rate} to
\begin{equation}\eqLabel{flow}
  \bv\flow\lr{\bv\radius_i} = \sum_{j=1}^{\numberOfParticles} \;\mat{M}_{ij}^\mathrm{TR}\cdot\bv\torque_j\text{.}
\end{equation}
In total, the resulting work then reads
\begin{equation}\eqLabel{work}
  \work\lr{t} = \int_0^{t} \sum_{i=1}^{\numberOfParticles} \bv\force_i\lr{ \cLR{\bv\radius}, t } \cdot \sum_{j=1}^{\numberOfParticles}\mat{M}_{ij}^\mathrm{TR}\cdot\bv\torque_j \;\Diff t'\text{,}
\end{equation}
and the heat reads
\begin{equation}\eqLabel{heat}
  \heat\lr{t} = \int_0^{t} \sum_{i=1}^{\numberOfParticles} \bv\force_i\lr{ \cLR{\bv\radius}, t } \cdot \LR{ \partialD{\bv\radius_i}{t} - \sum_{j=1}^{\numberOfParticles}\mat{M}_{ij}^\mathrm{TR}\cdot\bv\torque_j } \;\Diff t'\text{.}
\end{equation}
Note that, in contrast to systems driven by a linear shear flow, the work rate does \emph{not} trivially reduce to the virial expression for the configurational shear stress, as reported in previous studies \cite{Gerloff2018}.
All integrals and derivatives are evaluated using the Stratonovich calculus.
\subsection{Magnetic Torque Dependency}
\begin{figure}
  \centering\includegraphics[width=0.95\linewidth]{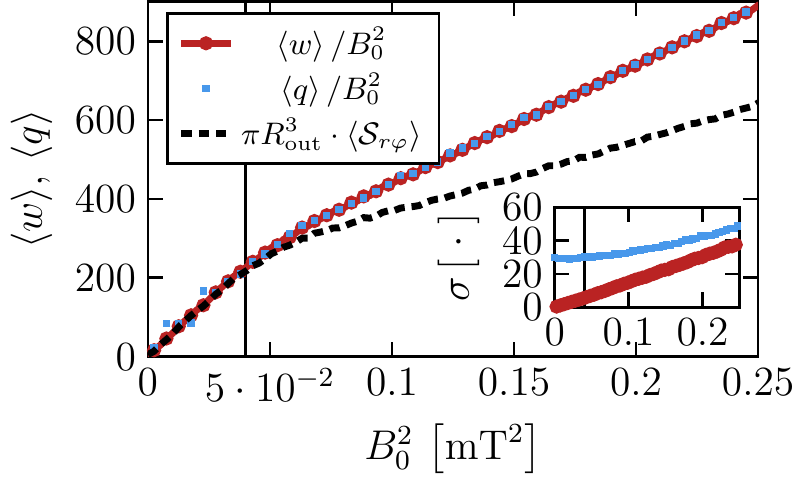}
  \caption{Mean work $\work$ (red circle) and heat $\heat$ (blue square) for integration time $t=0.1\,\brownianTime$ rescaled by the magnetic torque $\magneticField_0^2$. The rescaled mean shear stress is plotted for reference. The inset shows the standard deviation of the work and heat, respectively.}
  \figLabel{mean work and heat}
\end{figure}
In our numerical investigation, starting from the steady states obtained from the magnetic torque sweeps, plotted in \fref{mean angular velocity}(b), we compute work- $\work\lr{t}$ and heat trajectories $\heat\lr{t}$ for $t=0.1\,\brownianTime$ for the $N_\mathrm{ens}=10000$ systems to study the short-time behavior.
The resulting mean work and heat at $t=0.1\,\brownianTime$, as well as the corresponding standard deviation, is plotted in \fref{mean work and heat}.

Applying the magnetic torque $\magneticField_0^2$, $\qd{\work}$ displays a quadratic increase as a function of the torque $\magneticField_0^2$.
For supercritical magnetic torques, the work crosses over to another quadratic regime where $\qd{\work}$ increases slower as a function of $\magneticField_0^2$.
That is, the depinning transition at $\magneticField_{0,\mathrm{c}}^2$ is clearly reflected by the mean work as a function of $\magneticField_0^2$.
Interestingly, the same does not hold for the standard deviation of the work $\standardDeviation\LR{\work}$, plotted in the inset in \fref{mean work and heat}, which displays a linear increase with constant slope for all considered $\magneticField_0^2$.
The corresponding distributions are approximately Gaussian, yielding $\stochasticMoment_3\LR{\work} \approx 0$ and $\stochasticMoment_4\LR{\work} \approx 3$ for all $\magneticField_0^2$,

Turning now to the heat, we find that on \emph{average} the heat and the work are the same for all considered magnetic torques, as expected in a steady state.
However, the heat distributions display a finite width already in equilibrium, which remains approximately constant for subcritical magnetic torques $\magneticField_0^2 < \magneticField_{0,\mathrm{c}}^2$ and a subsequent linear increase for supercritical values $\magneticField_0^2 > \magneticField_{0,\mathrm{c}}^2$, as plotted in the bottom right inset in \fref{mean work and heat}.
In general, $\probability\lr{\heat}$ deviate from a Gaussian distribution, displaying a small positive skewness $\stochasticMoment_3\LR{\heat} \approx 0.2$ as well as a larger kurtosis $\stochasticMoment_4\LR{\heat} \approx 4$ for all $\magneticField_0^2$.

In \fref{mean work and heat}, we have plotted the mean work divided by the magnetic torque $\magneticField_0^2$, which allows for an easy comparison with the shear stress.
This comparison is motivated by our results for a planar slitpore system, where these two quantities are closely related \cite{Gerloff2018}.
However, the same does not hold true for the sheared circular system, which deviates from the expected relation for $\magneticField_0 > \magneticField_{0,\mathrm{c}}$.
\subsection{Time Dependency}
\begin{figure}
  \centering\includegraphics[width=0.95\linewidth]{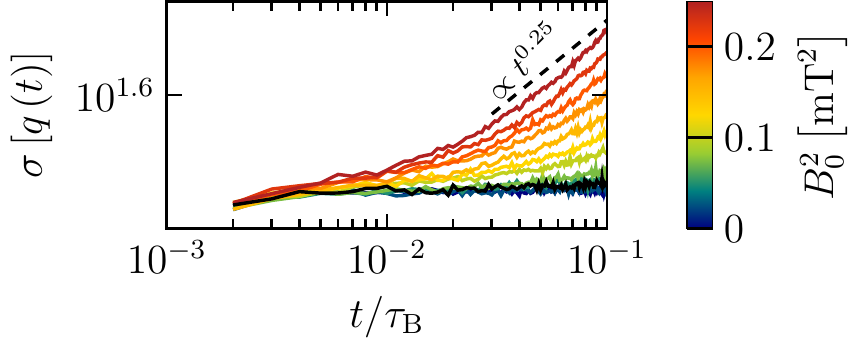}
  \caption{ Standard deviation of the heat as a function of time $t$ for various different magnetic torques. For reference, we have plotted a dashed line corresponding to $f\lr{t}\propto t^{0.25}$. The standard deviation at the critical magnetic torque is plotted in black.}
  \figLabel{heat standard deviation time evolution}
\end{figure}
As discussed in our earlier study of a planar slitpore system \cite{Gerloff2018}, the work and heat distributions are in general time dependent and the same behavior occurs here.
Regarding the mean values, we find a linear increase in time for both the work and the heat, that is, $\qd{\work\lr{t}} \approx \qd{\dot\work} t$ and $\qd{\heat\lr{t}} \approx \qd{\dot\heat} t$, and the functions are fully determined by their respective mean rates, as expected for a steady state.
Turning now to the time evolution of the standard deviation, we find a power law behavior for the work $\standardDeviation\LR{\work\lr{t}}\propto t^{0.55}$ for all considered magnetic torques.
Note that this is very close to $\standardDeviation\LR{\work}\propto \sqrt{t}$ that one finds for a single particle in a translated harmonic trap, yielding $\standardDeviation\LR{\work} = \sqrt{2\boltzmannConstant\temperature\qd{\work}}$ as follows directly from the integrated fluctuation theorem \cite{Seifert2012}.
In contrast, $\standardDeviation\LR{\heat\lr{t}}$ displays a more complex behavior, shown in \fref{heat standard deviation time evolution}, where the strength of the fluctuations saturate for intermediate times before transitioning to another power law behavior.
This saturation leads to a plateau, whose width seems to be connected to the typical waiting time of the particles as well as the plateau of the MSD discussed in \fref{mean squared angular displacement}.
However, a detailed investigation would require longer runs and better statistics.\\

Overall, we find that the depinning transition at $\magneticField_{0,\mathrm{c}}^2$ is clearly reflected by the mean values of the work and heat, as well as the strength of the heat fluctuations $\standardDeviation\LR{\heat}$ both with respect to the magnetic torque dependency and with respect to the time dependency.
A deeper interpretation of these results remains, at present, difficult due to the absence of analytical results for these types of strongly correlated systems.
\section{Conclusion}
Performing video microscopy experiments as well as Stokesian dynamics simulations, we have studied a dense, bidisperse colloidal system confined to a two-dimensional, disk-like cluster that is actuated by an external magnetic field.
The outer ring of particles are confined by harmonic traps and is kept static by time shared optical tweezers, whereas the inner ring, consisting of paramagnetic particles, is driven by a rotating magnetic field.
Focusing on rather small magnetic torques, we find a pronounced depinning transition, that is reminiscent of that occurring in incommensurate driven monolayers at finite temperature.

The dynamics at subcritical magnetic torques is characterized by a small, linear increase of the mean angular velocity per ring (with the applied torque), which stems from thermally activated slip-events.
During the latter, the locking of the inner rings to the static outer ring is momentarily broken and the inner rings slide past each other.
In this state, we observe both forward and backward slip events, where the probability of forward slips increases with increasing magnetic torque, yielding a net motion along the positive azimuthal direction.
At a critical magnetic torque, the system enters a second steady state, where the probability to find backward slip-events vanishes.
For this state, we find a bistability with respect to the mean angular velocity of the inner ring of the individual realizations.
This bistability is clearly reflected by a bimodal distribution both in simulations and experiments.

We can understand the bimodal distributions by analyzing the particle trajectories, from which we compute the waiting times between the slips as well as the jump times, corresponding to the duration of a slip.
For magnetic torques in the bistable region, the typical waiting time is of the order of the duration of the employed time average.
Thus, some realizations are momentarily locked, whereas others have performed a slip, corresponding to the slow and fast state of the bistability, respectively.
One interesting observation is that, for the first steady state, the typical waiting times of the different rings are approximately equal.
In contrast, for the second steady state the typical waiting time increases with the radial distance from the center.
This observation holds true for the jump times for all magnetic torques considered.

These typical waiting times are also reflected by the time-dependence of the mean squared displacement, which displays a pronounced plateau at intermediate times.
The corresponding range of times decreases with increasing the magnetic torque.
Such a sub-diffusive region is very common in strongly correlated sheared systems, such as colloidal glasses and dense liquid crystalline mixtures.
It is accompanied by an overshoot of the shear stress relaxation curves, which we also find for supercritical magnetic torques.
Both phenomena are connected to the breaking of particle cages, i.e. the plastic deformations during the slip events.
Overall, by monitoring the stress components, we find a pronounced shear thinning behavior at the critical magnetic torque, which is accompanied by a marked increase of the azimuthal- and radial pressure.
The latter corresponds to a radial expansion of the outer rings, which are observed both in simulation and experiments, as well as an compression of the inner ring.

Finally, we have briefly discussed the consequences of the depinning transition on two important stochastic thermodynamics quantities, i.e. the work and heat.
We find that the depinning transition is reflected by the mean work and heat as function of torque.
Moreover, we find signatures in the magnitude of the heat fluctuations as function of torque as well as in dependence the of integration time.
Interestingly, we did not observe a direct correspondence between the work rate and the shear stress, as we have reported for a planar slitpore system \cite{Gerloff2018}.\\

Overall, we find a very good agreement between numerical simulations and experiments not only with respect to the mean values but also the fluctuations of the mean angular velocity as well as the azimuthal- and radial displacements.
One open question is the importance of the magnetic torque dependency of the dipole-dipole interactions exerted between the paramagnetic particles.
Another interesting avenue is to analyze, in more detail, the plastic events during slips and their time- and space correlations.
Here, one goal is to predict the emergence of slips.
Finally, in the future, we aim to develop a deeper understanding of the stochastic thermodynamics in these strongly correlated systems.
One major challenge is that many exact results from stochastic thermodynamics make predictions for the entropy production, whose calculation is not trivial.
Research in this direction is in progress.
\section*{Conflicts of interest}
There are no conflicts to declare.
%
\appendix
\section*{Appendix}
\section{Hydrodynamic Interactions}\label{sec:hydrodynamics}
\begin{figure}
  \centering\includegraphics[width=0.95\linewidth]{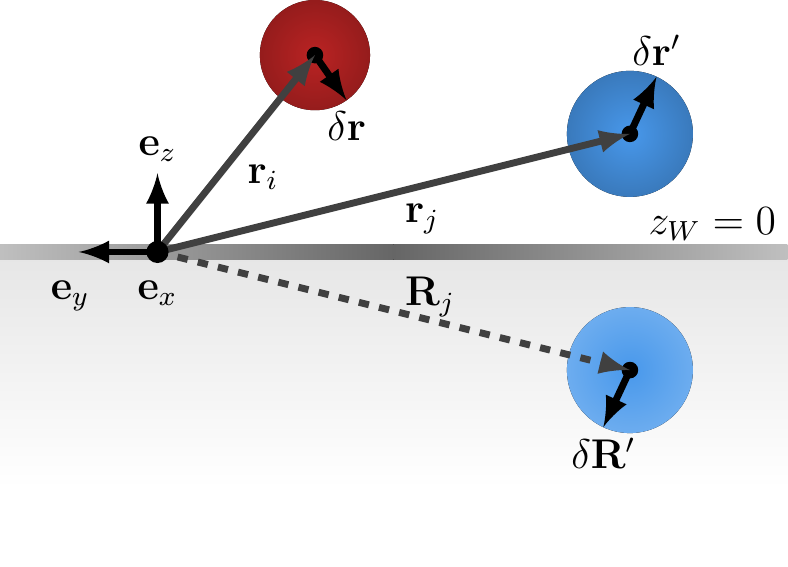}
  \caption{Sketch of the considered setup, with position $\bv\radius_i$, $\bv\radius_j$ of particle $i$ and $j$ as well as the image $\bv{R}_j$ of particle $j$ with respect to a plane boundary condition at $z_\mathrm{W}=0$.}
  \figLabel{blake sketch}
\end{figure}
The hydrodynamic interactions in our Stokesian dynamics simulations are modeled via the two mobility tensors $\mat{M}^\mathrm{TT}_{ij}$ and $\mat{M}^\mathrm{TR}_{ij}$ that encode the hydrodynamic flow acting on particle $i$ that is exerted from the translation and rotation of particle $j$, respectively.
The setup we have in mind consists of two spherical particles with different diameters close to a plane boundary located at $z_\mathrm{W} = 0$, as sketched in \fref{blake sketch}.

In the present study, we derive appropriate mobility tensors starting with the Blake solution \cite{Blake1971}
\begin{align}
      \blakeTensor\lr{\bv\radius_i,\bv\radius_j} &= \oseenTensor\lr{\bv\radius_i-\bv\radius_j} -\oseenTensor\lr{\bv\radius_i-\bv{R}_j} \nonumber\\
      \eqLabel{blake solution}
      &+ \unityMatrix^{z}\cdot\LR{2 z_j \mat{G}^\mathrm{D}\lr{\bv\radius_i-\bv{R}_j} - 2 z_j^2\mat{G}^\mathrm{SD}\lr{\bv\radius_i-\bv{R}_j} } \text{,}
\end{align}
where the first term is $\displaystyle \oseenTensor\lr{\bv\radius} = \frac{1}{8\pi\viscosity}\frac{1}{\radius}\LR{ \unityMatrix + \frac{\bv\radius\otimes\bv\radius}{\radius^2} }$ is the Oseen tensor between the two point particles at position $\bv\radius_i$ and $\bv\radius_j$, whereas the remaining terms represent the corrections from the plane boundary condition.
Further, $\displaystyle {\bv{R}_j = \bv\radius_j - 2 z_j \unitVector{z}}$ is the  position of the image of particle $j$, $\displaystyle {\unityMatrix^{z}=\unityMatrix-2\unitVector{z}\otimes\unitVector{z}}$ is a unit matrix with negative $zz$-component,
$\mat{G}^\mathrm{D}$ is a Stokes-doublet, and
$\mat{G}^\mathrm{SD}$ is a Source-doublet, which are defined as follows
\begin{equation}\eqLabel{stokes doublet}
  \mat{G}^\mathrm{D}\lr{\bv\radius_i-\bv{R}_j}=\nabla_{\bv{R}_j}\LR{ \oseenTensor\lr{\bv\radius_i-\bv{R}_j}\cdot\unitVector{z} }\text{,}
\end{equation}
and
\begin{equation}\eqLabel{source doublet}
\mat{G}^\mathrm{SD}\lr{\bv\radius_i-\bv{R}_j}= \frac{1}{2} \nabla^2_{\bv{R}_j} \oseenTensor\lr{\bv\radius_i-\bv{R}_j}
\end{equation}
with $\nabla_{\bv{R}_j}$ being the gradient with respect to to $\bv{R}_j$.
To account for the finite extend of the spherical particles we expand the Blake solution using the Faxén theorem \cite{Swan2007}
\begin{align}
  \eqLabel{faxen translation translation tensor}
   \mat{M}^\mathrm{TT}_{ij} &= \LR{1+\frac{\diameter_i^2}{24}\nabla_{\bv\radius_i}^2}\LR{1+\frac{\diameter_j^2}{24}\nabla_{\bv\radius_j}^2} \mat{G}^\mathrm{B}\lr{\bv\radius_i,\bv\radius_j}\\
   \eqLabel{faxen translation rotation tensor}
   \mat{M}^\mathrm{TR}_{ij} &=  \LR{1+\frac{d_i^2}{24}\nabla_{\bv\radius_i}^2}\frac{1}{2}\nabla_{\bv\radius_j}\times\mat{G}^\mathrm{B}\lr{\bv\radius_i,\bv\radius_j}\text{.}
\end{align}\\
Focusing first on $\mat{M}^\mathrm{TT}_{ij}$, plugging \eref{blake solution} into \eref{faxen translation translation tensor}, we find
\begin{align}
  \label{eq:translation translation self mobility}
  \mat{M}^\mathrm{TT}_{ii} &= \frac{1}{3\pi\viscosity\diameter_i}\unityMatrix - \mat{G}^\mathrm{RP}\lr{ \bv{R}_{ii} } + \delta\mat{G}\lr{\bv{R}_{ii}}\\
  \label{eq:translation translation mobility}
  \mat{M}^\mathrm{TT}_{ij} &= \mat{G}^\mathrm{RP}\lr{\bv\radius_{ij}} - \mat{G}^\mathrm{RP}\lr{ \bv{R}_{ij} } + \delta\mat{G}\lr{\bv{R}_{ij}}\text{,}
\end{align}
where $\displaystyle \mat{G}^\mathrm{RP}$ is the Rotne-Prager tensor and $\displaystyle \delta\mat{G}$ is the correction term from in addition to the Rotne-Prager contribution of the image particle, given by
\begin{widetext}
\begin{align}
  \mat{G}^\mathrm{RP}\lr{\bv\radius_{ij}} =& \frac{1}{6\pi \viscosity a_{ij}}\LR{\frac{3a_{ij}}{4\radius_{ij}}\lr{\unityMatrix+\frac{\bv\radius_{ij}\otimes\bv\radius_{ij}}{\radius_{ij}^2} } + \frac{a_{ij}^3}{2\radius_{ij}^3}\lr{\unityMatrix-3\frac{\bv\radius_{ij}\otimes\bv\radius_{ij}}{\radius_{ij}^2}} }\\
  \delta\mat{G}\lr{\bv{R}_{ij}} =& \frac{1}{8\pi\viscosity} \unityMatrix^{z}\cdot\LR{
  \frac{ \diameter_i^2\diameter_j^2}{12 R_{ij}^5}\unitVector{z}\otimes\unitVector{z} + \lr{ \frac{2 z_j}{R_{ij}^3} + \frac{ z_j \lr{ \diameter_i^2-\diameter_j^2 }}{2 R_{ij}^5}-\frac{5 \diameter_i^2 \diameter_j^2 Z_{ij}}{12 R_{ij}^7} }\unitVector{z}\otimes\bv{R}_{ij}\right.\nonumber\\
  &\left.+\lr{ -\frac{2 z_j}{R_{ij}^3} + \frac{ \diameter_j^2 Z_{ij}}{R_{ij}^5} + \frac{ z_j \lr{ \diameter_i^2-\diameter_j^2 }}{2R_{ij}^5}-\frac{5 \diameter_i^2 \diameter_j^2 Z_{ij}}{12 R_{ij}^7}}\bv{R}_{ij}\otimes\unitVector{z}\right.\nonumber\\
  &\left.+\lr{ \frac{6 z_i z_j}{R_{ij}^5}  - \frac{10\diameter_i^2\diameter_j^2}{48 R_{ij}^7} - \frac{10 Z_{ij} \lr{ z_i\diameter_j^2 +z_j\diameter_i^2} }{4 R_{ij}^7} + \frac{70 \diameter_i^2\diameter_j^2 Z_{ij}^2 }{48 R_{ij}^9} }\bv{R}_{ij}\otimes\bv{R}_{ij}\right.\nonumber\\
  &\left.+\lr{ -\frac{ 2 z_i z_j}{R_{ij}^3} - \frac{ \lr{\diameter_i^2-\diameter_j^2}Z_{ij}z_i }{2 R_{ij}^5} +\frac{ Z_{ij}^2\diameter_i^2}{2 R_{ij}^5} + \frac{\diameter_i^2\diameter_j^2}{24 R_{ij}^5} - \frac{ 10\diameter_i^2 \diameter_j^2 Z_{ij}^2 }{48 R_{ij}^7}  }\unityMatrix \;
 }\text{,}
\end{align}
\end{widetext}
with $a_{ij}=\sqrt{ \lr{d_i^2+d_j^2}/8 }$ being the effective radius of the two particles and $\bv{R}_{ij}=\bv\radius_i-\bv{R}_j$ being the distance between particle $i$ and the image of particle $j$.
Note that these expressions differ from that for polydisperse systems near a plane boundary condition, such as reported by Karzar-Jeddi et al. in Ref. \cite{Karzar-Jeddi2018}.
Unfortunately, the latter seem to contain errors as the reported expressions do not reduce to that of Ref. \cite{Swan2007} in the limit of $\diameter_i = \diameter_j$.
For a detailed derivation of the above mentioned expressions see Ref. \cite{Gerloff2020}.

Turning now to $\mat{M}^\mathrm{TR}_{ij}$, plugging \eref{blake solution} into \eref{faxen translation rotation tensor}, we find
\begin{align}\eqLabel{translation rotation self mobility}
  \mat{M}^\mathrm{TR}_{ii} =& 0\\
  \eqLabel{translation rotation mobility}
  \mat{M}^\mathrm{TR}_{ij} =& \frac{1}{8\pi\viscosity}\LR{ \frac{\bv\radius_{ij}}{\radius_{ij}^3} - \frac{ \bv{R}_{ij} }{R_{ij}^3} }\leviCivitaTensor\text{,}
\end{align}
where $\leviCivitaTensor$ is the Levi-Civita tensor, a third rank tensor which represents the cross product as follows ${\leviCivitaTensor\cdot\lr{\bv{A}\otimes\bv{B}}=\bv{A}\times\bv{B}}$.
Note that these expressions are identical to that of the monodisperse case reported in Ref. \cite{Swan2007}.
\section*{Acknowledgements}
This research was funded by the Deutsche Forschungsgemeinschaft (DFG, German Research Foundation) - Projektnummer 163436311 - SFB 910.

\clearpage


%


\end{document}